\documentclass[11pt]{article}
\newcommand{\be}{\begin{equation}}
\newcommand{\ee}{\end{equation}}
\newcommand{\beq}{\begin{eqnarray}}
\newcommand{\eeq}{\end{eqnarray}}

\def\theequation{\arabic{section}.\arabic{equation}}
\usepackage{graphicx}

\begin{document}
\begin{center}
{\bf \LARGE Quantum Tachyon Dynamics}\\
[7mm]
H. M. FRIED
\\
{\em Department of Physics \\
Brown University \\
Providence R.I. 02912 USA}\\
[5mm]
Y. GABELLINI
\\
{\em Institut Non Lin\'eaire de Nice\\
UMR 6618 CNRS\\
 1361 Route des Lucioles\\
06560 Valbonne France}\\
[5mm]

\vspace{5mm}
Abstract
\end{center}

It is suggested that charged tachyons of extremely large mass $M$ could not only contribute to the dark matter needed to fit astrophysical observations, but could also provide an explanation for gamma-ray bursts and ultra high energy cosmic rays. The present paper defines a quantum field theory of tachyons, the latter similar to ordinary leptons, but with momenta larger than energy. The theory is invariant under the full $CPT$ transformation, but separately violates $P$ and $T$ invariance. Lorentz invariance is broken for space-time intervals smaller than $M^{-1}$, but is effectively preserved for larger separations.

Topics treated include the choice and Schwinger Action Principle variations of an appropriate Lagrangian, spinorial wave functions, relevant Green's functions, a functional description of an $S$--Matrix and generating functional, and a variety of interesting kinematical processes, including photon emission and reabsorption, and relevant annihilation and scattering effects. A version of Ehrenfest's theorem is developed, in order to provide a foundation for a classical description of charged tachyons in an external elecromagnetic field. Applications are then made to three outstanding astrophysical puzzles : dark matter, gamma ray bursts and ultra high energy cosmic rays.
\newpage
{\bf\section{Introduction}}
\setcounter{equation}{0}
There are several astrophysical puzzles of high current interest -- dark matter, the source and content of the vacuum energy needed for an accelerating universe, the origins of gamma ray bursts and ultra high energy cosmic rays coming from galactic distances -- which may find a partial solution under the assumption that massive, charged tachyons exist in, and in the outer reaches of, each galaxy. The origin of tachyon pairs can be imagined in the immediate aftermath of the Big Bang, or more locally, by any cataclysmic event, such as a supernova explosion or black hole formation; they are definitely suggested by pair production in the intense, electromagnetic vacuum field fluctuations appearing in a `` bootstrap '' solution at extremely short distances \cite{one}. In this note, a brief QFT of tachyons as `` misgenerated '' leptons is presented, along with a sketch of their interactions with themselves, with photons, and with ordinary matter, in order to illustrate the possible relevance of such charged tachyons to the experimental puzzles stated above.

Two immediate theoretical observations can be made to set the stage for this discussion. The Lagrangians typically used to define the Higgs mechanism, and in particular the Weinberg -- Salam method of defining a `` true '' ground state, are basically theories of ``~condensed '' tachyons, theories which use a negative (mass)$^2$ and a self interaction to break spontaneously the initial vacuum symmetry, and arrive at a minimum energy ground state. But if a strong, external, electromagnetic field is locally present, and if the quanta of the tachyon field are charged, that ground state need not be, locally, the correct vacuum state; and one might expect fluctuations about that W--S vacuum. Those fluctuations could well be tachyonic, as expected from the form of the Lagrangian used, suggesting that pairs of charged tachyons could, in principle and just as for ordinary leptonic pairs, be produced by extremely strong electromagnetic fields.

Secondly, consider the non perturbative Schwinger mechanism \cite{two} for the production of lepton pairs by a strong electric field, a calculation which has received during the past half century a great amount of attention and generalization \cite{three}. Physically, the vacuum's charged lepton pairs which continuously fluctuate into and out of existence are acted upon by the continuous distribution of virtual photons comprising the external electric field; and if and when sufficient amounts of energy $E$ and momentum $p$ have been transferred to each member of a pair, they become real leptons, each with $E>p$. Suppose now that charged, tachyonic degrees of freedom exist, perhaps the same ones whose `` condensed '' summation defines the spontaneously broken vacuum. Then it is entirely reasonable to imagine that extremely strong electric fields could transfer sufficient energy and momentum to the tachyon pair, with each member here displaying $p>E$. There is no conservation law that forbids this; in principle, it is possible.

Of course, one immediate objection to $v>c$ motion is a resulting lack of micro--causality. But if the tachyon mass $M$ which sets the scale for acausal and Lorentz invariance violating effects is sufficiently large, e.g. $M\sim 10^7$ GeV, this difficulty will not be particularly relevant at typical laboratory distances. Nevertheless, the actions of charged tachyons after they are produced may well be unusual, especially at galactic distances.

Mention should be made of the (Wigner) classification \cite{four} of a relativistic field theory, wherein it is well known that a scalar tachyon field need have but one component, but that `` higher spin '' tachyons are represented by non compact groups (e.g. $SU(1,1)$ instead of $SU(2)$), and could have an infinite number of components. Nevertheless, our model field theory represents its leptonic tachyons in terms of four components, although somewhat modified from those of Dirac. We also insist on the restriction to orthochronous Lorentz transformations (LTs), so that -- in contrast to the work of ref\cite{five} -- we absolutely forbid LTs which can interchange positive and negative energy tachyons. An amount of causality sufficient to define scattering states requires a special assumption, but can be arranged. Invariance under the full $CPT$ operation is maintained, although $P$ and $T$ invariance are separately violated. And we stress that the Lorentz invariance (LI) of this model field theory is broken at very small space--time intervals. But if massive, charged tachyons provide a partial explanation of how the large scale universe works, so be it.

Several decades ago, other authors \cite{five} have discussed the possibility of zero charge tachyons, and produced somewhat simpler models than the leptonic tachyons suggested here. Following the spirit of ref\cite{five}, it would be possible to describe spin zero charged tachyons, which couple in a standard, gauge invariant manner to the photon field, with an interaction term of the form :
\be ieA_{\mu}\bigl(\varphi^{\dagger}\partial_{\mu}\varphi - \partial_{\mu}\varphi^{\dagger}\varphi ) + e^2\varphi^{\dagger}A^2\varphi\ee
where $A_{\mu}$ denotes a conventional photon field operator. What makes this theory unsuitable for our astrophysical purposes is the $e^2\varphi^{\dagger}A^2\varphi$ term of this interaction, which can describe the simultaneous emission of a pair of photons; in this case the Cerenkov--like angular restrictions of single photon emission and absorption are absent, which restrictions are present when but a single photon is emitted by a tachyon. Crucial to the astrophysical usefulness of the present model tachyons is the strong probability that a very high energy tachyon can reabsorb every photon that it emits, and thus retain its high energy across astrophysical times and distances. But if two photons were emitted simultaneously, the probability of reabsorption of the photon moving in a direction roughly opposite to that of the tachyon would be small.

The arrangements of these remarks is as follows. In the next Section, a `` misgenerated lepton '' model of tachyons is defined, and its spinorial content outlined. In Section 3, a complete, functional Quantum Tachyon Dynamics ( QTD ) is set up, with coupling to the electromagnetic field, in analogy to QED. In the next Section, the possibility of photon emission and reabsorption by charged tachyons is discussed, and in Section 5 the kinematics of tachyon particle reactions is outlined. A version of Ehrenfest's theorem is exhibited in Section 6, to set the stage for a subsequent `` Classical Tachyon Dynamics '', describing the classical motion of a charged tachyon in an external electromagnetic field. A final Summary and an Appendix complete the presentation.
\bigskip
{\bf\section{QTD as a QFT}}
\setcounter{equation}{0}
The QFT here called QTD is modeled as closely as possible upon QED, with but two changes, and one change of interpretation. The standard QED Lagrangian may be written as :
\be{\cal L}\,=-\,\bar\psi\,\Bigl[\, m + \gamma\!\cdot\!(\partial - ieA)\,\Bigr]\psi + {\cal L}_0(A)\ee
where ${\cal L}_0(A)$ denotes the free photon Lagrangian, and $m$ and $e$ are the electron bare mass and charge, respectively; in all of the following, hermitian $\gamma_{\mu}$ satisfying $\{\gamma_{\mu}, \gamma_{\nu}\}= 2\delta_{\mu\nu}$ will be used, with $\{\gamma_{5}, \gamma_{\mu}\}= 0$,  $\gamma_j=-i\gamma_4\alpha_j$, $\alpha_j=\sigma_1\otimes\sigma_j$, $\gamma_5=\gamma_1\gamma_2\gamma_3\gamma_4$, where the elements of $\gamma_{4,5}$ are two--by--two matrices : $\gamma_4=\pmatrix{1&0\cr0&-1\cr}$, $\gamma_5=-\pmatrix{0&1\cr1&0\cr}$. The $\sigma_j$ denote the hermitian Pauli matrices, satisfying $\sigma_i\sigma_j=\delta_{ij}+i\epsilon_{ijk}\sigma_k$. We use the Minkowski metric~: $x_{\mu}=(\vec x, ict)$, with $\hbar=c=1$ and $\bar\psi=\psi^{\dagger}\gamma_4$.

From (2.1) and the Schwinger Action Principle, one obtains both the field equations
\be\Bigl[\, m + \gamma\!\cdot\!(\partial - ieA)\,\Bigr]\psi\,=0\ee
 and, from the surface--term variations of the Action integral of (2.1), the equal time anti commutation relation ( ETAR ) :
\be \Bigl\{\psi_{\alpha}(x), \bar\psi_{\beta}(y)\Bigr\}_{x_0=y_0}\!\!\!\!\!\!\!=\gamma_4^{\alpha\beta}\,\delta(\vec x - \vec y)\ee
 with (2.3) providing a statement of micro--causality, as expected in a causal theory.

The first, and simplest, departure from QED that shall be adopted here is the replacement of $m$ by $iM$, where $M$ is understood to be large, $M\gg m$~: 
\be{\cal L}\rightarrow-\,\bar\psi_{T}\,\Bigl[\, iM + \gamma\!\cdot\!(\partial - ieA)\,\Bigr]\psi_T + {\cal L}_0(A)\ee
This is not the final tachyon Lagrangian, but only the simplest, initial form written to display the needed $m^2\rightarrow -M^2$ continuation. For simplicity, we allow the tachyon field to be coupled to the electromagnetic field with the QED coupling constant, and first consider the properties of the free tachyonic field, $\psi_T^{(0)}$, satisfying :
\be(\,iM + \gamma\!\cdot\!\partial\,)\,\psi_T^{(0)}\,=0\,=\bar\psi_T^{(0)}(\,iM - \gamma\!\cdot\!\overleftarrow\partial\,)\ee
In direct imitation of the free lepton fields' representations in term of creation and destruction operators, and spinorial wave functions, we write :
\be\displaystyle\psi_T^{(0)}(x)=(2\pi)^{-3/2}\sum_{s=1}^2\int \!\!d^3\!p\left[{iM\over E(p)}\right]^{1/2}\,\left\{b_s(\vec p)\,e\,^{\displaystyle ip\!\cdot\!x}u_s(p) + d_s^{\dagger}(\vec p)\,e\,^{\displaystyle -ip\!\cdot\!x}v_s(p)\right\}\ee
where $b_s(\vec p)$ and $d_s^{\dagger}(\vec p)$ are the destruction and creation operators, respectively, for $T$ and $\bar T$, while $u_s(p)$ and $v_s(p)$ are their appropriate spinors; and $p\!\cdot\!x=\vec p\!\cdot\!\vec x - E(p)x_0$, with $E(p)=\sqrt{\vec p^2 - M^2}$. Perhaps the simplest description appears if $b_s(p)$ and $d_s(p)$ satisfy exactly the same anticommutation rules as for ordinary leptons. One then realizes that the $\exp\left(\pm ix_0\sqrt{\vec p^2 - M^2}\right)$ factors of (2.6) will generate exponentially growing or damping time dependence of the $\psi_T^{(0)}$, $\bar\psi_T^{(0)}$, which is simply not acceptable. The obvious way to prevent this is to restrict $\vert\vec p\vert$ values to be larger than $M$, a solution which will have other consequences, but ones that are acceptable and reasonable in a non causal theory. Henceforth, a factor of $\theta(\vert\vec p\vert-M)$ will appear under the momentum integrals of $\psi_T^{(0)}$ and $\bar\psi_T^{(0)}$.

Consider now the momentum--space equations which the tachyon $T$ and anti--tachyon $\bar T$ must satisfy :
\be(\,M + \gamma\!\cdot\!p\,)\,u_s(p)\,=0\,,\ \ \ \ \ \ (\,M - \gamma\!\cdot\!p\,)\,v_s(p)\,=0 \ee
Solutions of (2.7) may be expressed in terms of the projection operators $\Lambda_{\pm}(p)=\displaystyle{M\pm\gamma\!\cdot\!p\over2M}$~:
\be u_s(p)\,=a\Lambda_-(p)\,\xi_s\,,\ \ \ \ \ \ v_s(p)\,=a\Lambda_+(p)\,\xi_{s+2} \ee
where, as for leptons, $s=1$ or 2, and $a$ is an appropriate normalization constant. The adjoint spinors $\bar u=u^{\dagger}\gamma_4$, $\bar v=v^{\dagger}\gamma_4$, are then given by :
\be \bar u_s(p)\,=a^{*}\,\bar\xi_s\,\Lambda_+(p)\,,\ \ \ \ \ \ \bar v_s(p)\,=a^{*}\,\bar\xi_{s+2}\,\Lambda_-(p) \ee
and one sees that the norms of these states have the unfortunate property of vanishing : $\bar u_s(p)\,u_s(p)=\bar v_s(p)\,v_s(p)=0$.

Does this signify that it is impossible to construct lepton--style tachyonic states ? Not necessarily, because, -- as Dirac pointed out, long ago --  one can define expectation values with the aid of an indefinite metric operator, $\eta=\eta^{\dagger}$, such that the proper norm of a state is given by $\bar u\eta u$, rather than $\bar uu$. In our case, there exists a clear candidate for this metric operator : $\eta=\gamma_5$, which has the pleasant property of converting $\Lambda_{\pm}$ to $\Lambda_{\mp}$ as it passes through either operator. However, there is a price to pay for this convenience, in that such norms are no longer scalars but pseudoscalars; and the vectors constructed from them will be pseudovectors, etc, while expected vectors may be replaced by vectors and axial vectors. It is difficult to avoid the suspicion that such tachyons may turn out to be associated with the electroweak interactions, forming the `` condensate '' of the W--S vacuum, although that subject will not be treated in this paper.

Following the above discussion, we shall insist that a factor of $\gamma_5$ be inserted immediately after each $\bar u_s$ or $\bar v_s$ factor, for all matrix elements subsequently calculated. With this intention, the tachyonic part of the Lagrangian of (2.4) shall now be written as :
\be{\cal L}_T\,=-\,\bar\psi_{T}\,\gamma_5\,\Bigl[\, iM + \gamma\!\cdot\!(\partial - ieA)\,\Bigr]\psi_T \ee
with field equations :
\be\Bigl[\, iM + \gamma\!\cdot\!(\partial - ieA)\,\Bigr]\psi_T \,=0\,=\bar\psi_T\gamma_5\,\Bigl[\, iM - \gamma\!\cdot\!(\overleftarrow\partial + ieA)\,\Bigr]\ee
so that the $\gamma_5$ insertion is automatic. Were the spinors $\xi_s$ chosen as for the leptons, $\xi_s=\chi_s$, where $\chi_1$ is the column matrix of elements $(1, 0, 0, 0)$, $\chi_2$ has elements $(0, 1, 0, 0)$, etc, then one easily calculates :
\be \bar u_{s'}(p)\gamma_5u_s(p)\,={\vert a\vert^2\over2M}\,\chi_{s'}^{\dagger}\vec\sigma\!\cdot\!\vec p\,\chi_s\ee
which, while non zero, does not display the usual $\delta_{ss'}$, but corresponds to a linear combination of the spin states labeled by $s$ and $s'$. With $\hat p=\vec p/\vert\vec p\vert$, this difficulty can be removed by the choice of the modified spinors $\xi_s\rightarrow\xi_s(\hat p)$ :
$$\xi_1\,=\pmatrix{U_1\cr0\cr0\cr0},\ \ \ \xi_2\,=\pmatrix{0\cr U_2\cr0\cr0},\ \ \ \xi_3\,=\pmatrix{0\cr0\cr U_1\cr0},\ \ \ \xi_4\,=\pmatrix{0\cr0\cr0\cr U_2}$$
and $U_1(\hat p)=\exp\left[ i\vec\sigma\!\cdot\!(\hat p\times\hat e_3)\right]$, $U_2(\hat p)=\exp\left[ -i\vec\sigma\!\cdot\!(\hat p\times\hat e_3)\right]$. If the normalization is chosen as~: $\vert a\vert=\displaystyle\sqrt{2M\over\vert\vec p\vert}$, the combination :
$$ \bar u_{s'}(p)\gamma_5u_s(p)\,=U_{s'}^{\dagger}(\hat p)(\vec\sigma\!\cdot\!\vec p)U_s(\hat p)\,=i\delta_{ss'}$$
and similary : $ \bar v_{s'}(p)\gamma_5v_s(p)\,=-\,i\,\delta_{ss'}$.

The closure, or completeness relation for such spinors is then given by :
\be\sum_{s=1}^2\left\{u_s^{\alpha}(\bar u_s\gamma_5)^{\beta} - v_s^{\alpha}(\bar v_s\gamma_5)^{\beta}\right\}\,=i\,\delta_{\alpha\beta} \ee
The free field spinors now satisfy :
$$(\,M + \gamma\!\cdot\!p\,)\,u_s(p)\,= (\,M - \gamma\!\cdot\!p\,)\,v_s(p)\,=0$$
and :
$$\bar u_s(p)\gamma_5(\,M + \gamma\!\cdot\!p\,)\,=\bar v_s(p)\gamma_5(\,M - \gamma\!\cdot\!p\,)\,=0$$
It should be noted that the Action formed from the Lagrangian of (2.10) is hermitian, as is the Hamiltonian. These assertions require the customary neglect of spatial surface terms, as well as the observation -- obtain from the field equations -- that : $\displaystyle{\partial\over\partial t}\int\!\! d^3x\,\bar\psi_T\,\gamma_5\gamma_4\psi_T\,=0$.

If the coefficient of $A_{\mu}$ in (2.10) is defined as the charged tachyon current operator : $J_{\mu}^T = ie\bar\psi_T\,\gamma_5\gamma_{\mu}\psi_T$, then it follows from the field equations (2.11) that this charged current is conserved : $\partial_{\mu}J_{\mu}^T=0$. Charged conjugation may be effected by the same set of Dirac matrices used for the lepton case : $\psi_c=C\bar\psi^T$, where the superscript $T$ denotes transposition and $C=i\gamma_2\gamma_4$.

Finally, one must discuss the equal time anticommutation relation $ \Bigl\{\psi_T^{\alpha}(x), \bar\psi_T^{\beta}(y)\Bigr\}_{x_0=y_0}$, which -- as for ordinary leptons -- are assumed to be the same for free and interacting field operators. The Schwinger Action Principle \cite{six} underlying QFT specifies the ETAR by identifying the infinitesimal generators from the surface terms of the variation of the Action~: $\delta W=\displaystyle-\!\int\!\! d\sigma_{\mu}\bar\psi\gamma_{\mu}\delta\psi$ in QED, or : $\delta W=\displaystyle-\!\int\!\! d\sigma_{\mu}\bar\psi_T\gamma_5\gamma_{\mu}\delta\psi_T$ for the present case of QTD. For the usual case where the relevant space--time surface is taken as a flat time cut, one has in QED :
$$\delta W=\displaystyle-\!\int\!\! d\sigma_{4}\bar\psi\gamma_{\mu}\delta\psi=i\!\int\!\! d^3y\,\bar\psi(\vec y, t)\gamma_{\mu}\delta\psi(\vec y, t)$$
thereby identifying the field momentum operator conjugate to $\psi_{\alpha}$ as $\pi_{\alpha}(\vec y, t)=\left(\bar\psi(\vec y, t)\gamma_4\right)_{\alpha}$ and from this follows the choice of ETAR : $\Bigl\{\psi_{\alpha}(x), \psi_{\beta}^{\dagger}(y)\Bigr\}_{x_0=y_0}\!\!\!\!\!\!\!=\delta_{\alpha\beta}\,\delta(\vec x - \vec y)$, which expresses the fermion micro--causality of conventional QED.

For an acausal theory, one should not expect to require strict micro--causality; but, keeping to the original statement of the Action Principle, it is possible to obtain a modified expression of the QTD ETAR, which can be written in the form :
\be\Bigl\{\psi_T^{\alpha}(x), \left(\bar\psi_T(y)\gamma_5\right)^{\beta}\Bigr\}_{x_0=y_0}\!\!\!\!\!\!\!=(\gamma_4)^{\alpha\beta}\,\hat\delta(\vec x - \vec y)\ee
It is then straightforward to employ (2.13) and show that the substitution of (2.6) and its adjoint into the LHS of (2.14) yields :
\be\hat\delta(\vec x - \vec y)=\int\!\!{d^3p\over(2\pi)^3}\,\theta(\vert\vec p\vert - M)\,e\,^{\displaystyle \vec p\!\cdot\!(\vec x - \vec y)}\ee
which is just the form one expects with the restriction $\vert\vec p\vert>M
$. This `` modified delta function~'' of (2.15) is a well defined distribution of  $\vert\vec x - \vec y\vert$, which can be written in various form, such as :
\be\hat\delta(\vec x - \vec y)=\delta(\vec x - \vec y) - \int\!\!{d^3p\over(2\pi)^3}\,\theta(M - \vert\vec p\vert)\,e\,^{\displaystyle \vec p\!\cdot\!(\vec x - \vec y)}\ee
where the second RHS of (2.16) is a finite measure of the lack of micro--causality of this tachyon theory. With $\vec z=\vert\vec x - \vec y\vert$, it is immediately evaluated as : 
\be{4\pi M\over z^2}\left\{{\sin(Mz)\over Mz} - \cos(Mz)\right\}\ee
For small and large $Mz$, (2.17) reduces to $\displaystyle{4\pi\over3}M^3$, $Mz\ll 1$; and to $-\displaystyle{4\pi M\over z^2}\cos(Mz)$, $Mz\gg 1$. In other words, the lack of micro--causality occurs mainly for extremely small distances, $z<M^{-1}$, where it is significant, while, for $z>M^{-1}$, the effect is oscillatory and of unimportant magnitude. The basic theory, however, is non causal in a qualitatively different sence, with tachyon propagation emphasized outside the light cone but damped away inside, in exactly the opposite sense to that of QED.

The free tachyon propagator, defined as :
\beq{}&\displaystyle S_T^{\alpha\beta}(x-y)=i<\left(\psi_T^{\alpha}(x)\left(\bar\psi_T(y)\gamma_5\right)^{\beta}\right)_+>\nonumber\\&\equiv\displaystyle i\theta(x_0-y_0)\!<\psi_T^{\alpha}(x)\left(\bar\psi_T(y)\gamma_5\right)^{\beta}\!>-\,i\theta(y_0-x_0)\!<\left(\bar\psi_T(y)\gamma_5\right)^{\beta}\psi_T^{\alpha}(x)\!>\eeq
is then, with (2.14), to satisfy :
\beq{}&\displaystyle (iM+\gamma\!\cdot\!\partial)\,S_T(x-y)=\delta(x_0-y_0)<\left\{\left(\gamma_4\psi_T(x)\right)^{\alpha}, \left(\bar\psi_T(y)\gamma_5\right)^{\beta}\right\}>\nonumber\\&\displaystyle =\delta(x_0-y_0)\,\hat\delta(x-y)\nonumber\eeq
and has the integral representation :
\be S_T(z)=-i\!\!\int\!\!{d^4p\over(4\pi)^4}\,\theta(\vert\vec p\vert - M){e\,^{\displaystyle ip\!\cdot\!z}\over(M+\gamma\!\cdot\!p\,)}\ee
An immediate question is the method of choosing contours to perform the momentum integrals of (2.19), since : $(M+\gamma\!\cdot\!p\,)^{-1}=(M-\gamma\!\cdot\!p\,)/(M^2-p^2)^{-1}$. If, in the usual way, $M^2\rightarrow M^2\pm i\epsilon$, one must decide which possibility is appropriate. Calculation of the $\displaystyle\int\!\!dp_0$ yields :
\be S_T(z)=(\mp){1\over2}\int\!\!{d^3p\over(2\pi)^3}\,{\theta(\vert\vec p\vert - M)\over\sqrt{\vec p^2-M^2}}\,\,e\,^{\displaystyle i\vec p\!\cdot\!\vec z\mp\vert z_0\vert\sqrt{\vec p^2-M^2}}\!\left( M - \vec\gamma\!\cdot\!\vec p\pm i\gamma_4\epsilon(z_0)\right)\ee
and if the $z\rightarrow0$ limit of (2.20) is compared with a direct computation of the second line of (2.18), using the completeness statement (2.13), the two calculations will agree if the upper sign is used in (2.20); that is : $M^2\rightarrow M^2+i\epsilon$. For $z\ne0$, one then finds agreement with the form of the fermion propagator in QED, in the sense that an exponential factor $\exp( i\vec p\cdot\vec z-i\omega\,\vert z_0\vert)$ is present in both theories, with $\omega=\sqrt{\vec p^2+m^2}$ for QED and $\omega=\sqrt{\vec p^2-M^2}$ for QTD. Henceforth, $M^2\rightarrow M^2+i\epsilon$.

Because of the $\theta(\vert\vec p\vert - M)$ factor, the propagator of  (2.19) and (2.20) is not Lorentz invariant (LI), but one expects the lack of such invariance to be significant only for $\vert\vec z\vert<M^{-1}>\vert z_0\vert$, and unimportant for $\vert\vec z\vert>M^{-1}<\vert z_0\vert$. The easiest way to see this is to replace $\theta(\vert\vec p\vert - M)$ by $1-\theta(M-\vert\vec p\vert)$, and in analogy to the ordinary lepton case, set $S_T(z)=(iM- \gamma\!\cdot\!\partial)\,\Delta_T(z)$, so that $\Delta_T$, and therefore $S_T$, is given by the difference of the two terms :
\be\Delta_T(z)=\Delta_c(z,m^2)\left\vert_{m^2\rightarrow-M^2}\right.-\int\!\!{d^4p\over(2\pi)^4}\,\theta(M-\vert\vec p\vert){e\,^{\displaystyle ip\!\cdot\!z}\over p^2-M^2-i\epsilon}\ee
The first RHS term of (2.21) is just the LI, scalar propagator function of $M^2z^2$ obtained from the ordinary $\Delta_c(z,m^2)$ by interchanging its properties inside and outside the light cone. The second RHS term contains the LI--violating dependence, which can be displayed as : 
\be {M\over2\pi^2z_0}\left[{e\,^{\displaystyle-i\pi/4}\over2}\,{\sin(M\vert\vec z\vert)\over M\vert \vec z\vert}-\phi(M\vert z_0\vert)\right]\ee
with $\phi(u)=\displaystyle\int_0^1{x\,dx\over\sqrt{1-x^2}}\,\,e\,^{\displaystyle-xu}$, a smoothly varying, finite function of $u$ which can be given in terms of Bessel and Struve functions. But its limits for small and large $u$ are easily obtained :
$$\phi(u)\left\vert_{u\ll1}\right.\sim1+O(u)\,,\ \ \ \phi(u)\left\vert_{u\gg1}\right.\sim{1\over2u^2}\left(1+O(1/u^2)\right)$$
which illustrates the remark above concerning the unimportance of the LI--violating terms for $\vert\vec z\vert$ and $\vert z_0\vert>M^{-1}$. One expects that the significant Physics of tachyons will involve distances and time intervals much larger than $M^{-1}$, so that such violations of LI will be as unimportant as they are unmeasurable.
\bigskip
{\bf\section{Functional QTD}}
\setcounter{equation}{0}
In this section, a generating functional for QTD is constructed, in analogy with that of ordinary QFT, with charged tachyons replacing charged leptons. Because the properties of the free tachyon creation and destruction operators have been chosen to be the same as those of ordinary lepton fields, the transition from $n$--point functions of QTD to corresponding $S$--Matrix elements can be taken over directly from ordinary QED. However, because of the ubiquitous $\theta(\vert\vec p\vert - M)$ factors, one requires the understanding that tachyons $IN$ and $OUT$ field operators must be defined at `` asymptotic times '' that are outside the light cone of any origin of coordinates in terms of which the field operators might be evaluated.

For the Lagrangian :
\be {\cal L}\,=-\,\bar\psi_T\gamma_5\,\Bigl[ iM + \gamma\!\cdot\!\left(\partial - ie[A+A^{ext}]\right)\,\Bigr]\psi _T+ {\cal L}_0(A)\ee
where $A$ denotes the ordinary, quantized, photon field, and  ${\cal L}_0(A)$ is its free Lagrangian (~in an appropriate gauge ), one defines the generating functional in terms of tachyonic and photon sources as :
\be <S[A^{ext}]\!>{\cal Z}\left[\eta, \bar\eta, j\,\right]=<S[A^{ext}]\bigg(e\,^{\displaystyle i\int\left[\bar\eta\psi_T+\bar\psi_T\eta+j_{\mu}A_{\mu}\right]}\bigg)_+\!\!>\ee
just as one would do for QED. With the aid of the field equations, and the photon ETCR and the tachyonic ETAR -- or, directly, from the Schwinger Action Principle -- one obtains \cite{six} :
\beq{}&\displaystyle{\cal Z}\left[\eta, \bar\eta, j\,\right]=e\,^{\displaystyle {i\over 2}\int \!\!jD_cj}\!\cdot\,e\,^{\displaystyle -{i\over 2}\int\!\! {\delta\over\delta A}D_c{\delta\over\delta A}}\cdot\,e\,^{\displaystyle i\!\int\!\!\bar\eta\, G_T[A+A^{ext}]\gamma_5\,\eta}\nonumber\\&\cdot\,\,e\,^{\displaystyle L[A+A^{ext}]}/\!\!<S[A^{ext}]\!>\eeq
where, for (3.3), $A_{\mu}=\displaystyle\int \!\!d^4y D_{c,\mu\nu}(x-y)j_{\nu}(y)$. Because of the normalization requirement ${\cal Z}[0,0,0]=1$, there follows : 
\be<S[A^{ext}]\!>\,\,=e\,^{\displaystyle -{i\over 2}\int\!\! {\delta\over\delta A}D_c{\delta\over\delta A}}\cdot\,e\,^{\displaystyle L[A+A^{ext}]}\left\vert_{A\rightarrow 0}\right.\ee
where :
\be L[A]\,=i\int_0^e\!de'\,{\rm Tr}\left[\gamma\!\cdot\!A\,G_T[e'A]\right]\ee
and :
\be G_T[eA]\,=S_T\left[ 1 - ie(\gamma\!\cdot\!A)S_T\right]^{-1}\ee

One can here define `` retarded '' and `` advanced '' tachyonic Green's function :

$S_{R,A}=(iM-\gamma\!\cdot\!\partial)\Delta_{R,A}$, with :
\be\Delta_R(z)\,=\int\!\!{d^4p\over(2\pi)^4}\,{\theta(\vert\vec p\vert - M)\,{e\,^{\displaystyle ip\!\cdot\!z}}\over p^2-M^2-i\epsilon s(p_0)}\ee
where $s(p_0)=p_0/\vert p_0\vert$; a similar representation, but with the sign of $s(p_0)$ reversed, defines $\Delta_A(z)$.Both Green's functions satisfy : $(M^2+\partial^2)\Delta_{R,A}(z)=\delta(z_0)\hat\delta(\vec z)$, so that the $S_{R,A}$ satisfy : $(iM+\gamma\!\cdot\!\partial)S_{R,A}(z)=\delta(z_0)\hat\delta(\vec z)$. Both $\Delta_R$ and $S_R$ vanish for $z_0<0$, and hence merit the subscript $R$; and conversely for $\Delta_A$ and $S_A$, which vanish for $z_0>0$. Again, LI of these functions is not exact, but the violatins occur only for very small space--time distances. This formalism permits physical significance to be maintained for orthochronous Lorentz transformations, with a well defined sense of time and of a particle's sign of energy, while restricting consideration to Lorentz transformations that are performed outside the light cone.

From the above discussion, the initial step of Symanzik's derivation \cite{six} of the functional reduction formula, between every $S$--Matrix element and the generating functional, is valid. It is :
\be\psi_T(x)\,=\sqrt{Z_2}\,\psi_T^{IN}(x) + \int\!\!d^4y\,S_R(x-y)\,{\cal D}_y\,\psi_T(y)\ee
where ${\cal D}_y=iM+\gamma\!\cdot\!\partial_y$, $M$ is the renormalized ( physical ) tachyon mass, $\psi_T(x)$ denotes the fully interacting ( Heisenberg representation ) tachyon operator, and $\psi_T^{IN}$ is its free field counterpart, bearing its renormalized mass. $Z_2$ is the tachyon wave function renormalization constant. Since ${\cal D}_x\,\psi_T^{IN}=0$, and $\theta^2=\theta$, the operation of ${\cal D}_x$ on both sides of (3.8) yields a simple identity. Once the validity of (3.8) is appreciated, all of Symanzik's functional steps follow through, with the result :
\beq{}&\displaystyle S/\!\!<S[A^{ext}]\!>\,=\,:\,\exp \left\{ Z_3^{-1/2}\!\!\int\!\!A_{\mu}^{IN}(-\partial^2){\delta\over \delta j_{\mu}}\right.\nonumber\\&\displaystyle\left. +\, Z_2^{-1/2}\!\!\int\!\!\left[ \bar\psi_T^{IN}\gamma_5\,\vec{\cal D}{\delta\over \delta\bar\eta} - {\delta\over \delta \eta}\overleftarrow{\cal D}\gamma_5\,\psi_T^{IN}\right]\right\}:\,{\cal Z}\left[\eta,\bar\eta,j\right]\left\vert_{\eta,\bar\eta,j\rightarrow0}\right.\eeq
With (3.9), it is possible, in principle, to calculate the amplitude of any process to any perturbative order; in addition, and as in certain limiting situations in ordinary QFT, it may be possible to sum subsets of classes of Feynman graphs, with each class containing an infinite number of graphs.
\bigskip
{\bf\section{Photon Emission and Reabsorption}}
\setcounter{equation}{0}
There is one important difference between the Physics of certain `` modified bremsstrahlung~'' processes in QED and QTD which is worth mentioning. In conventional $S$--Matrix descriptions of reactions initiated by a few particles, the initial and final states are understood to be asymptotically well separated, and this is expected in a world where all massive particles travel with $v<c$. But when charged tachyons and photons are involved, the situation is not as clear, because physical `` overlaps '' of these particles can exist for macroscopic times; and the final, physical result can be quite unexpected.

As an example, consider the amplitude for a charged tachyon of momentum $T_{\mu}$ to emit a photon, and to leave the scene of the reaction with momentum $T_{\mu}'$. Kinematically, this process resembles Cerenkov radiation, with $T=T'+k$. Squaring and summing both sides of this 4--vector equation, with $T^2=T'^2=M^2$, $k^2=0$, leads to the restrictions : $T'\!\!\cdot\!k=\vec T'\!\!\cdot\!\vec k-T_0k_0=0$. In other words, the angle between the photon's spatial momentum $\vec k$, and the direction, $\hat T$, of the emitting tachyon, is given by $\cos\theta=T_0/\vert\vec T\vert$; and similarly, $\cos\theta'=T_0'/\vert\vec T'\vert$ defines the angle between the outgoing $\vec k$ and $\vec T'$. Note that  these kinematical relations are independant of the energy of the emitted photon. Using standard Rules, as sketched in the Appendix, one can calculate the probability per unit time for the emission of a photon of arbitrary polarization within a band of energy $\omega$ to $\omega + d\omega$, as :
\be {1\over T}\,{d\over d\omega}\,\sum\left\vert<T'\!,k\,\vert S\,\vert \,T\!>\,\right\vert^2\,={\alpha M^2\over \vert\vec T\vert T_0}\ee
where $\alpha=e^2/4\pi$. ( A differential cross--section is not appropriate here, since the angular distributions are fixed by the kinematics above ). With a physical upper cut--off chosen as $\omega_{max}=T_0$, the lowest order probability/time for such emission is :
\be{1\over T}\,\sum\left\vert<T'\!,k\,\vert S\,\vert \,T\!>\,\right\vert^2\,={\alpha M^2\over \vert\vec T\vert}\ee
By direct calculation, (4.2) has exactly the same probability/time for the inverse process, the absorption of a photon $k$ by a tachyon $T$ which results in $T'$. In order for this inverse process to become significant, the incident tachyon should move in a directed photon beam, at just the right kinematical angle, for an amount of time $\tau=\vert\vec T\vert/\alpha M^2$, using the perturbative result for the inverse process as indicative of that of a more precise calculation. Because the tachyon moves faster than $c$, the question to be posed is whether the outgoing photon of the emission process stays `` in close proximity '' to the outgoing tachyon for a duration of time of this magnitude, and can therefore be reabsorbed by that outgoing tachyon. The angles for emission and absorption for the two processes are identical, and therefore the reabsorption of that emitted photon is a kinematic possibility. The question to be answered is : how long are the outgoing photon and tachyon `` in close proximity '' ?

The relative velocity of emitted photon and final tachyon is the relevant quantity. With $c=1$ ( and $\hbar=1$ ), the velocity vector of the emitted photon is $\hat k$, while that of the final tachyon is $\vec v_T'=dT_0'/d\vec T'=\vec T'/T_0'$, with $T_0'=\sqrt{\vec T'^2-M^2}$. The relative velocity of the two is $\vec v_{rel}=\hat k-\vec T'/T_0'$. Because of the spatial, kinematical relations stated above, the projection of $\vec v_{rel}$ along the photon's direction vanishes : 
\be \hat k\!\cdot\vec v_{rel}\,=0\ee
Also, the projection of $\vec v_{rel}$ along the tachyon's direction is small : 
\be\hat T'\!\cdot\vec v_{rel}\,=\hat k\!\cdot\hat T' -{\vert \vec T'\vert\over T_0'}\,={T_0'\over\vert\vec T'\vert} - {\vert \vec T'\vert\over T_0'}\,=-{M^2\over T_0'\vert \vec T'\vert}   \ee
for a high energy tachyon, $T_0'\gg M$ ( for which case $T_0\gg M$ as well ). A better indication of their relative velocity might simply be 
\be\vec v_{rel}^2\,=\left(\hat k - {\vec T'\over T_0' }\right)^2\,=1 + {\vec T'^2\over T_0'^2} - 2{\hat k\!\cdot\vec T'\over T_0'}\,=+{M^2\over T_0'^2}\ee
again using the kinematic relation $\hat k\!\cdot\vec T'=T_0'$. During a time $\tau$ one would then expect the photon and tachyon to `` move apart '' by a distance $D\sim \tau(M/T_0')$. The significance of eqs. (4.3)--(4.5) is that for a high energy tachyon, with $T_0\gg M$ and $T_0'\gg M$, the emitted photon and final tachyon separate very slowly. Using (4.5) as a measure of their velocity of relative separation, and the time $\tau\sim \vert\vec T'\vert/\alpha M^2$ as a qualitative measure of the time suggested by (4.2) for the probability of reabsorption to be $\sim O(1)$, the distance of that separation grows to $D=v_{rel}\,\tau\sim{\vert\vec T'\vert\over T_0'}/\alpha M\sim 10^{-18}$ cm, for $M\sim 10^7$ GeV and $T_0'\gg M$. If a value larger than 1/137 is used for $\alpha$, or a larger value for $M$, this distance $D$ is even shorter.

Contrast this with the distance moved by the tachyon, of velocity $\vert\vec T'\vert/T_0'$, during the same interval $\tau$ : $D_{T'}=v_{T'}\,\tau\sim\vec T'^2/\alpha M^2T'_0$, which is a factor $\vert\vec T'\vert/M$ larger than $D$. The final, high energy tachyon therefore moves through a distance which is considerably larger than the separation distance between it and the emitted photon; and this suggests that the photon and final tachyon remain `` in close proximity '', such that during this time the photon is reabsorbed by that tachyon. At least, there should be a non zero probability of reabsorption; and continued in this way, over asymptotic times and distances, there should be a significant, non zero probability that a very high energy tachyon will reabsorb every photon that it emits, and thereby retain its original high energy. If not exactly, then almost so.

 Another argument leads to a similar conclusion. Consider the external $\vec E$ and $\vec B$ fields produced by a charged tachyon moving at constant velocity. Any external field can be represented as an integral over virtual photon modes, with weighting depending on the nature of the particular external field. If that field vanishes when viewed by an observer at the Cerenkov--like angle, it means that no photons are emitted at that angle. But emission at that angle is precisely the definition of a real photon, since then and only then can $k^2=0$; if the $\vec E$ and $\vec B$ fields of a constant velocity tachyon do vanish when viewed at this angle, such a classical result indicates that no real photons can be emitted by a tachyon in a Cerenkov--like process. A straightforward calculation of those fields shows that, if assumed continuous, they indeed must vanish, precisely and uniquely at that angle.\bigskip
{\bf\section{Kinematics of Tachyon--Particle Reactions}}
\setcounter{equation}{0}
The conjecture of the preceeding paragraph underlies the reactions of this Section, in which high energy tachyons are able to retain their energy/momenta while traveling to galactic distances, where they annihilate and/or scatter with photons and other, ordinary matter.

One does not have the freedom to make Lorentz transformations that `` simplify '' calculations by moving to a Lorentz frame in which the $T_0$ component of one tachyon vanishes, for such a transformation assigns that tachyon an infinite velocity. This does not make any physical sense; the mathematical equivalent of that step is dividing by zero.

Imagine that a charged $T\!\!-\!\bar T$ pair is produced at some space--time point, either by a Schwinger mechanism following from quantized QED vacuum fluctuations at extremely small distances \cite{one}, or by some other mechanism associated with the Big Bang, or subsequent, cataclysmic events. These tachyons immediately separate, each with $v>c$, moving away from the world of ordinary particles into the space between stars, and into the outer reaches of every galaxy. But galactic magnetic fields exist, which must influence these charged tachyons in the conventional way, by bending their trajectories into partially `` circular '' paths, depending on the direction and magnitude of the magnetic fields encountered. ( If there are  a sufficient number of such tachyons, they themselves can be thought of as generating at least a portion of the galactic magnetic fields; but that subject is reserved for a separate treatment ). The picture that emerges is of `` swarms '' of charged tachyons, moving not quite randomly in curved orbits at galactic distances; and every so often, depending on the density of such tachyons and of stray photons and bits of ordinary matter, there will occur collisions.
\vskip0.3truecm
A) $T\!\!-\!\bar T$ Annihilation

\hskip1truecm i) `` Fusion '' into a single, neutral particle : $T+\bar T=P$, with $P^2=-\mu^2$, $M\gg\mu$. This process can only happen if the angle between $\hat T$ and $\hat{\bar T}$ happens to agree with the $\theta(T_0,\bar T_0)$ obtained by squaring and summing both sides of the momentum/energy equation~:
\be\cos\theta(T_0,\bar T_0)\,={T_0\bar T_0-M^2-\mu^2/2\over\sqrt{M^2+T_0^2}\sqrt{M^2+\bar T_0^2}}\ee
There are several limits of (5.1) which are interesting : 

-- if $T_0$ and $\bar T_0\gg M$ : $\cos\theta\rightarrow 1$ ( a `` parallel '' annihilation );

-- if $0<(T_0\ {\rm and}\ \bar T_0)\ll M$ : $\cos\theta\ge -1$ ( an almost `` head--on '' annihilation );

-- if $T_0\gg M$ but $\bar T_0\ll M$ : $\cos\theta\rightarrow \bar T_0/M-M/T_0-\mu^2/(2MT_0)$, ( an `` interpolating '' annihilation ). 

In other words, when $T_0$ and $\bar T_0\gg M$, and their velocities are only slightly larger than $c$, a $T$ and a $\bar T$ can `` fuse '' together in a parallel collision, becoming a single ( neutral ) particle of $v<c$, and with enormous energy, $p_0=T_0+\bar T_0$. But when $T_0$ and $\bar T_0\ll M$, and their velocities are arbitrarly high, they can only annihilate into a single particle by an almost head--on collision. The general case is covered by (5.1), whose LHS must always lie between $-1$ and $+1$.

\hskip1truecm ii) Into photon pairs : $T+\bar T=k+k'$. Perhaps the simplest case to calculate is a head--on collision : $\vec T+\vec{\bar T}=0$, $\bar T_0=T_0$, with the outgoing photons moving away from each other at a relative angle of $\pi$, for which $k_0=k_0'=T_0$. Here is a possible mechanism for producing gamma bursts at galactic distances with no $X$--ray or optical tails.

\hskip1truecm iii) Into particle pairs : $T+\bar T=p+\bar p$. Again, assume a head--on collision of $T$ and $\bar T$, with the outgoing particles separating from each other at an angle of $\pi$; the result for $p_0$ and $\bar p_0$ is exactly the same as for annihilation into a pair of gammas, $p_0=\bar p_0=T_0$. Here is a possible mechanism for producing ultra high energy cosmic rays. Further, if $p$ and $\bar p$ are charged, they can radiate, opening other possibilites for galactic $X$--ray and $\gamma$--ray emissions.
\vskip0.3truecm
B) Scattering : $T+p=T'\!+p'$

\hskip1truecm Suppose that a particle ( e.g. a proton floating about in outer galactic space ) is struck by an energetic tachyon. We idealize this situation to a proton at rest struck by a tachyon of energy $T_0\gg m$, write the conservation laws in the form $T+p-p'=T'$, square and sum both sides to find an exact, linear equation ( assuming $p_0'>m$ ) :
\be p'_0\,={N\over D}\,m\,={(m+T_0)^2+(M^2+T_0^2)\cos^2\theta\over(m+T_0)^2-(M^2+T_0^2)\cos^2\theta}\,m\ee
where $\theta$ is the angle between $\vec T$ and $\vec p\,'$. Since $N>0$, one expects that the largest value of $p_0'$ will result from the smallest value of $D$, or the maximum possible value of $\cos^2\theta$. From energy conservation, $p_0+T_0-p_0'=T_0'>0$, which when applied to the present case of $p_0=m$, requires :
\be\cos^2\theta<{T_0(m+T_0)^2\over (M^2+T_0^2)(2m+T_0)}\,\equiv \cos^2\theta_{max}<1\ee
Substitution of the $\cos^2\theta_{max}$ of (5.3) into (5.2) then yields :
\be p_0'\,=m\,\left[{1+\displaystyle{1\over 1+2m/T_0}\over 1-\displaystyle{1\over 1+2m/T_0}}\right]\,\simeq T_0\left(1 - {m\over T_0}+\cdots\right)\ee
An effective `` billiard ball '' collision has occurred, with the final proton taking on the energy of the incidental tachyon.

This example shows that an initial tachyon is able, by scattering, to transfer a considerable amount of its energy to a particle initially at rest ( or moving at a modest $v<c$ ), and in the process create high energy cosmic rays. Radiation from the scattering of charged particles by this mechanism should, in principle, contribute $X$--ray and $\gamma$--ray backgrounds.\bigskip
{\bf\section{From Ehrenfest's Theorem to Loop Annihilation}}
\setcounter{equation}{0}
It has been suggested above that charged tachyons of very high energy could reabsorb any photon emitted, thereby preserving their four--momentum across galactic times and distances; and, as they move away from and between galaxies, such tachyons could be influenced by galactic magnetic fields ( of orders of magnitude from milli to micro Gauss ) and caused to move in `` circular '' paths, at least while they are under the influence of a coherent magnetic field.

A fundamental question arises as to just how a charged tachyon would be affected by specified electromagnetic fields. There are no `` classical '' tachyons in our world, upon which experiments could be performed to yield Classical Tachyon Dynamics ( CTD ), and so the analysis must proceed in the opposite sense : from the assumed QTD Lagrangian of (2.10) to a one--particle representation of that Lagrangian, and thence to a subsequent approximation recognized as representing the `` classical '' action of a specified electromagnetic field on a single, charged tachyon. These last steps may be realized by a derivation of the corresponding Ehrenfest's theorem, and that is the subject of this Section.

Begin first with the free tachyon field operator of (2.6), and consider its one--particle matrix element :
$$\phi(x)\,=<0\,\vert\,\psi_T(x)\,\vert\,\vec p,s\!>$$
which satisfies the same equation as does $\psi_T(x)$, and has its solution :
$$\phi(x)\,=N\,e\,^{\displaystyle ip\!\cdot\!x}\,u_s(p)$$
with $N$ an appropriate normalization constant. Because $u_s(p)$ satisfies : $(M+\gamma\!\cdot\!p)u_s(p)\,=0$, upon multiplication by $i\gamma_4$ one obtains :
$$( i\gamma_4M+\vec\alpha\!\cdot\!\vec p\,\, )\,u_s(p)\,=p_0\,u_s(p)$$
which permits the identification of $H_T^{(0)}=iM\gamma_4+\vec\alpha\!\cdot\!\vec p$ as the one--free--tachyon Hamiltonian operator, with real eigenvalue $p_0$. Note that this Hamiltonian is not hermitian, but its eigenvalue is real ( see ref\cite{eight} for examples of such Hamiltonians ). In QED, the same analysis provides the one--free--electron Hamitonian, $H^{(0)}=m\gamma_4+\vec\alpha\!\cdot\!\vec p$.

The gauge invariant extension of this one--tachyon Hamltonian to include the effects of an external electromagnetic field is immediate :
$$H_T\rightarrow iM\gamma_4+\vec\alpha\!\cdot\!( \vec p - e\vec A ) + eA_0$$
just as one does in QED.

The time dependence of the one--tachyon wave function, assumed properly normalized according as : $\int \!d^3x\,\phi^{\dagger}\gamma_5\phi=1$ is given by the relevant Schr\"odinger equation $i\hbar\,{\partial\phi/\partial t}=~H_T\phi$, and its adjoint $-i\hbar\,{\partial\phi^{\dagger}/\partial t}=\phi^{\dagger}H_T^{\dagger}$, while the expectation value of any tachyonic operator $Q$ is defined by :
$$<Q>\,=\int\!\!d^3x\,\phi^{\dagger}\gamma_5\,Q\,\phi$$
As usual in this `` Schr\"odinger representation '', the full time dependence is carried by the wave functions, and the operators $Q$ are understood to be time independent. But the $c$--number $A_{\mu}^{ext}(x)$ are allowed to depend upon time, since unitary transformations on $c$--numbers cannot effect their time dependence. Hence, the total time rate of change of $<Q>$ is given by :
$${d\!<Q>\over dt}\,=\int\!\!d^3x\,\phi^{\dagger}\gamma_5\,\left({\partial Q\over\partial t}+{1\over i\hbar}[\,Q,H_T]\right)\phi$$
where, since $\{ \,\gamma_5,\gamma_4\,\}=0=[ \,\gamma_5, \vec \alpha \,\,]$, $H_T=\gamma_5H_T^{\dagger}\gamma_5$. Thus, if $Q\rightarrow\vec r$, 
$${d\!<\vec r>\over dt}\,=\int\!\!d^3x\,\phi^{\dagger}\gamma_5\,\vec\alpha\,\phi$$
showing, as in QED, that $\vec\alpha$ is the operator whose expectation value corresponds to velocity.

The average momentum of a charged tachyon in an external electromagnetic field is calculated from the gauge invariant, `` canonical '' contribution : 
$$<\vec p>\,=\int\!\!d^3x\,\phi^{\dagger}\gamma_5\,( \vec p - e\vec A )\,\phi$$
and one easily finds that :
\beq{}\displaystyle {d\!<\vec p>\over dt}&=&\displaystyle-e\int\!\!d^3x\,\phi^{\dagger}\gamma_5\,{\partial\vec A\over\partial t}\,\phi + {1\over i\hbar}\int\!\!d^3x\,\phi^{\dagger}\gamma_5\,\left[(\vec p - e\vec A), H_T\right]\,\phi\nonumber\\&=&\displaystyle e\int\!\!d^3x\,\phi^{\dagger}\gamma_5\,\left[ -\vec\nabla A_0-{\partial\vec A\over\partial t}+ \vec\alpha\times\vec B\right]\,\phi\\&=&\displaystyle e\int\!\!d^3x\,\phi^{\dagger}\gamma_5\,\left[ \vec E + \vec\alpha\times\vec B\right]\,\phi\nonumber\eeq
Exactly the same relation, but without the $\gamma_5$ factor, is obtained in QED.

If we now suppose that $\vec E$ and $\vec B$ are spatially constant over the dimensions of the tachyon -- or, more precisely, over the wave packet that describe the tachyon, as it moves with $v>c$ -- then (6.1) can be rewritten as :
$${d\!<\vec p>\over dt}\,=e\left[\, \vec E +<\vec v>\!\times\vec B\,\right]$$
which is just the conventional relation for a charged particle moving in specified $\vec E$ and $\vec B$ fields, with $\vec p$ and $\vec v$ now replacing $<\vec p>$ and $<\vec v>$. If we wish to write the corresponding four vector as : 
$$p_{\mu}\,=M\,{d\!x_{\mu}\over d\tau}\,=Mu_{\mu}$$
where $\displaystyle\sum_{\mu}u_{\mu}^2=+1$ ( rather than the $-1$ of the Minkovski metric ) and :
$${dp_0\over d\vec p}\,={\vec p\over p_0}\,=\vec v$$
the covariant equation for the motion of a charged particle is again valid, as in QED : 
$${d^2x_{\mu}\over d\tau^2}\,={e\over M}\,F_{\mu\nu}(x){d\!x_{\nu}\over d\tau}$$
except that the tachyon mass $M$ replaces the electron's $m$, and $dt/d\tau=\gamma=1/\sqrt{v^2-1}$.

With this demonstration that the motion of a charged tachyon follows essentially the same classical equations as those of other charged particles, it is now possible to understand how such charged tachyons can contribute to dark matter. Consider the production of a burst of charged $T$s, and $\bar T$s, by some catastrophic process. They separate, presumably moving in roughly opposite directions; and, concentrating on the $T$s, assume they possess a distribution of velocities moving roughly in the same direction. The lead $T$ of this group, moving with the fastest velocity, generates a magnetic field which it outruns and cannot feel; but the next $T$ behind it will be affected by that field, and deflected so that it eventually lines up behind the first $T$. And the same process happens all way down the line, resulting in a `` line '', or effective current of such charged $T$s.

Imagine that this current line is moving trough galactic space -- the space between galaxies -- where there exist magnetic fields of magnitude $B\sim 10^{-6}$ Gauss, which are coherent across distances $R\sim 10^4$ light years. One naturally expects such $T$s to fall into orbits defined ( reinstating the $c$ dependence ) by :
\be Mv^2/R\,=\gamma^{-1}evB/c\ee
where $\gamma^{-1}=[(v/c)^2-1]^{1/2}$, and $v>c$. For high energy $T$s, $v\ge c$, and so let us choose $(v/c)^2=1+10^{-6}$, $\gamma=10^3$, and insert the observed $B$ and $R$ values to obtain from (6.2) an estimate of what we might expect for $M$. Amusingly, the result is $M\equiv 10^7$ GeV, which is just the order of magnitude found in the original calculation of ref.\cite{one}.

The point of this argument is that if those $B$ fields arise from currents in the visible galactic matter, then the charged $T$s trapped in orbits by those $B$s are `` rigidly '' connected to the matter that we can see; but the $T$s themselves are invisible, they are `` dark ''. (~Note that were any coherent cyclotron radiation possible by such loops, its typical frequency would be far too small to detect, $\omega=v/R\simeq 10^{-14}$ s$^{-1}$ ). If there are enough such charged $T$s, they could contribute to the dark, or missing matter of current observation. This argument can be easily generalized to include $T$s of all energies.

Further, if there exist two or more such loops, of roughly the same size, parallel to each other and separated horizontally by a distance on the order of their radius, then their own magnetic fields would combine with that of the $B$ which hold them to create a `` magnetic bottle '' effect, trapping any available number of ordinary charged particles, which would then also contribute to dark matter. Two such loops, each with their `` macro--macroscopic~'' magnetic moments pointing along the magnetic field $\vec B$ would form a stable system : their magnetic moments attract but their like electric charges repel each other, leading to an equilibrium separation distance of order of their loop radius.

Finally, if one imagines that one loop consists of $\bar T$ circulating in the opposite direction to that of the $T$ loop, their loop magnetic moments point in the same direction along the line separating them, and the $T$ and $\bar T$ loops would be attracted towards each other -- both their magnetic moments and like  charges attract --  eventually leading to a cataclysmic `` loop annihilation '', which might even be visable astronomically. If the $T$ and $\bar T$ loops encircle a large gas cloud, as they well might, then the `` perfect fusion~'', loop--annihilation could emit sufficient radiation and newly formed high velocity matter to irradiate and compress the gas, setting off internal shock waves, and perhaps even a subsequent nuclear reaction. The corresponding $\gamma$--ray production, with $X$--ray and optical tails, and a natural afterglow, could possibly match the well known, observed GRB data \cite{nine}.\bigskip
{\bf\section{Summary}}
\setcounter{equation}{0}
In the belief that any model which can simultaneously relate astrophysical/experimental  results involving vacuum energy, dark matter, and the origin of ultra high energy cosmic rays and gamma bursts is worth considering, we have here put forth the idea of charged tachyons, as a possible, underlying mechanism. We note in passing the intriguing possibility of charged tachyons as responsible, in part, for observed, galactic magnetic fields, but provide in this paper no further discussion. Rather, we hope that the QTD analysis sketched above will generate further studies in this direction.

Among those topics which it would be useful to understand in clear detail, are the following.

1) Possible radiation signatures from the scattering of tachyons on ordinary particles, at galactic distances, especially if the tachyon charge is different from the charge of the particle it strikes.

2) Induced emission and absorption of CMB photons.

3) The use of these charged tachyons as a possible condensation--mechanism for the Weinberg--Salam vacuum.

4) The incorporation of such tachyons into a General Relativistic framework.

One final remark. One may feel uncomfortable without a proper definition of the tachyon vacuum state, in terms of conventional particle theory. Let us ask the familiar question : what is the mass of a quark ? The well known answer is that it is physically impossible to measure directly its mass, its intrinsic energy; and that one must be content with estimates that can be inferred from measurements which display results that depend on the type of measurement. If quantum mechanics teaches us anything, it is that no statements should be made -- or can be believed -- about quantities that cannot be measured. But we infer that a quark must have an intinsic mass, and we assume that a quark vacuum state exists.

In an analogous way, one cannot directly measure any properties of our assumed tachyons, but they can be used to explain, or to contribute to the understanding of certain, puzzling, astrophysical observations. Does our lack of ability for direct measurements mean that there exists no tachyon vacuum ? Not at all; it merely reflects the fact that it is impossible for us to measure directly the properties of an isolated, individual tachyon, just as for the quark situation, but for different kinematical and dynamical reasons. Again, all we can do is to infer the properties of tachyons by matching their predictions against astrophysical observations; and, again, we can only assume the existence of a tachyon vacuum.
\vskip0.5truecm
{\bf Acknowledgments}
\vskip0.3truecm
The authors warmly thanks Prof. Ian Dell'Antonio for several, most informative conversations, Dr. Walter Becker for pointing us to ref\cite{four} and Dr. Fr\'ed\'eric Daigne for detailed informations on GRB observations.

\vskip1cm
\def\theequation{A.\arabic{equation}}
{\bf\section*{Appendix }}
\setcounter{equation}{0}
\appendix
A brief sketch of the probability/time `` intrinsic bremsstrahlung '' production, to lowest perturbative order, eq. (4.1), proceeds as follows.

The $S$--matrix element of the lowest order, free field, photon and tachyon operators :
\be<p',k\,|\,S\,|\,p\!>\,=ie\!\int\!\!d^4x\!<p',k\,|\,\bar\psi_T(x)\gamma_5\gamma\!\cdot\!A(x)\psi_T(x)\,|\,p\!>\ee
may be calculated by inserting into (A.1) the conventional photon operator, and the tachyon field operator (2.6) and its adjoint, which yields :
\be<p',k\,|\,S\,|\,p\!>\,={ie\over\sqrt{2\pi}}\,\sqrt{M^2\over EE'}{1\over\sqrt{2\omega}}\left(\bar u_{s'}(p')\gamma_5\gamma\!\cdot\!\epsilon\, u_s(p)\right)\delta^{(4)}(p-p'-k)\ee
where $\epsilon_{\mu}$ is the photon polarization, and we use $\vec p$ and $E(p)=\sqrt{\vec p^2-M^2}$ to represent the tachyon coordinate. Upon calculating $|\!\!<p',k\,|\,S\,|\,p\!>\!|^2$, the square of  $\delta^{(4)}(p-p'-k)$ is, as always, replaced by $(2\pi)^{-4}VT\,\delta^{(4)}(p-p'-k)$. The relevant quantity to calculate is then not $|\!\!<p',k\,|\,S\,|\,p\!>\!|^2$ summed over all $\vec p$ and $\vec k$ values and polarizations ( and averaged over initial and summed over final spin indices ), but the probability for this process to occur per unit time :
\beq{}\displaystyle {1\over T}\sum|\!\!<p',k\,|\,S\,|\,p\!>\!|^2\,&=&\displaystyle{M^2e^2\over(2\pi)^2E}\int\!\!{d^3k\over2\omega}\,{\delta(E(\vec p)-E(\vec p-\vec k)-\omega)\over E(\vec p)-\omega}\nonumber\\&\displaystyle&\times\sum_{s,s'}{1\over 2}\,\vert \,\bar u_{s'}(p-k)\gamma_5\gamma\!\cdot\!\epsilon\, u_s(p)\,\vert^2\eeq
The spin and polarization sums produce exactly unity, and the continuous energy $\delta$ function of (A.3) may be replaced by $(E-\omega)/(\omega p)\,\delta(\cos\theta-E/p)$ where $\theta$ is the angle -- fixed by the conservation laws -- between $\hat p$ and $\hat k$. The result is :
$${\alpha M^2\over pE}\int_0^{\omega_{max}}\!\!d\omega$$
and we choose the maximum possible value of $\omega_{max}$ as $E$, yielding the result quoted in Section 4.


\vskip1truecm
\begin{thebibliography}{**}

\bibitem{one} H.M. Fried, arXiv: hep-th/0310095 (9 October 2003). 
\bibitem{two} J. Schwinger, Phys. Rev. {\bf82}, 664 (1951).
\bibitem{three} See, for example, M.N. Hounkonnou and M. Naciri, J. Phys. G{\bf 26}, 1849 (2000). More recent attempts at incorporating arbitrarily varying electromagnetic fields can be found in the papers of T.N. Tomaras, N.C. Tsamis and R.P. Woodard, Phys. Rev. D{\bf62}, 125005 (2000); H.M. Fried, R.P. Woodard, Phys. Lett. B{\bf524}, 233  (2002); J. Avan, H.M. Fried and Y. Gabellini, Phys. Rev. D{\bf67}, 16003 (2003).
\bibitem{four} J. Beckers and M. Jaspers, Ann. Phys. {\bf113}, No. 2, 237 (1978).
\bibitem{five} J. Dhar and E.C.G. Sudarshan,  Phys. Rev. {\bf174}, No. 5, 1808 (1968).
\bibitem{six} See, for example, H.M. Fried, `` Functional Methods and Models in Quantum Field Theory '' , The MIT Press, 1972.
\bibitem{seven} K. Symanzik, UCLA Lectures, reproduced in ref.\cite{six} above.
\bibitem{eight} C. M. Bender and C. Boettchner, Phys. Rev. Lett. {\bf80}, 5243 (1998); C. M. Bender, D. C. Brody and H.F. Jones, Phys. Rev. Lett. {\bf89}, 270401 (2002) and {\bf93}, 251601 (2004).
\bibitem{nine} F. Daigne, `` Gamma Ray Bursts '', EDP Sciences, Springer--Verlag (1999); T. Piran, Rev. Mod. Phys. {\bf76}, 1143 (2004); F. Frontera, arXiv: astro-ph/0407633. 
\end{thebibliography}
\end{document}